\documentclass[aps,prl,twocolumn,superscriptaddress,showkeys,floatfix]{revtex4}

\usepackage[colorlinks=true,citecolor=blue,linkcolor=blue]{hyperref}
\usepackage{graphicx}
\usepackage{latexsym}
\usepackage{amsmath,amssymb}
\usepackage{lipsum}
\usepackage{xcolor}
\usepackage[utf8]{inputenc}
\usepackage[english]{babel}
\usepackage{lineno}
\usepackage{float}
\DeclareUnicodeCharacter{2061}{}
\usepackage{xspace}
\usepackage{comment}
\usepackage{xr}
\addto\captionsenglish{}

\makeatletter
\let\@fnsymbol\@fnsymbol@latex
\@booleanfalse\altaffilletter@sw
\makeatother

\newcommand{\editor}[2]{%
  \expandafter\newcommand\csname #1note\endcsname[1]{%
    \textcolor{#2}{[\textbf{#1:}  ##1]}}%
  \expandafter\newcommand\csname #1\endcsname[1]{%
    \textcolor{#2}{##1}}%
  \expandafter\newcommand\csname #1cancel\endcsname[1]{%
    \textcolor{#2}{\sout{##1}}}%
  \expandafter\newcommand\csname #1change\endcsname[2]{%
    \textcolor{#2}{\sout{##1} ##2}}%
  \newenvironment{#1text}{\color{#2}}{\color{black}}
}

\begin{document}
\graphicspath{{figures/}}

\title{Tuning competing electronic phases in monolayer VSe\texorpdfstring{$_2$}{2} via interface hybridization}

\author{Ishita Pushkarna\textsuperscript{*}}
\author{Árpád Pásztor\textsuperscript{*}}

\affiliation{Department of Quantum Matter Physics, University of Geneva, 1211 Geneva, Switzerland}

\author{Greta Lupi}
\author{Adolfo O. Fumega}
\affiliation{Department of Applied Physics, Aalto University, 02150 Espoo, Finland}

\author{Christoph Renner}
\affiliation{Department of Quantum Matter Physics, University of Geneva, 1211 Geneva, Switzerland}


\begin{abstract}
     Competing electronic phases in two-dimensional transition metal dichalcogenides constitute a fertile platform for uncovering emergent ground states and elucidating the control parameters that govern the correlated electron phases. Among these materials, vanadium diselenide is particularly compelling: while the bulk hosts a well-established charge density wave (CDW), monolayers exhibit markedly different electronic behavior. Here, we identify three distinct electronic regimes in mechanically exfoliated VSe$_2$ flakes on Au(111) substrates, where interfacial hybridization, charge transfer, and strain act as primary tuning parameters of electronic order. Monolayers strongly coupled to gold show complete suppression of the CDW, accompanied by the emergence of moiré modulations. In contrast, bilayers preserve the in-plane $4a \times 4a$ CDW characteristic of the bulk limit. Strained, electronically decoupled monolayers formed in suspended membrane and bubble regions stabilize a $\sqrt{3}a\times\sqrt{7}a$ CDW phase, underscoring the reversible role of substrate interaction and hybridization.
\end{abstract}

\maketitle
%


Competing electronic phases and their tunability are of central interest in contemporary solid-state physics. Traditional tuning parameters include chemical doping, hydrostatic pressure, strain, and magnetic field. With the emergence of two-dimensional (2D) materials, new possibilities appear, such as dimensionality \cite{Pásztor_2017_2dmat, hwang_large-gap_2022}, electrostatic gating \cite{Costanzo_2016, wu_electrostatic_2023, jo_electrostatically_2015}, and controlled stacking into heterostructures enabling proximity effects, moiré physics, and interface engineering \cite{Pushkarna_2023, Sun_2023, huang_emergent_2020}. VSe$_2$ is particularly interesting in this context. It is a metallic transition-metal dichalcogenide (TMD) which, in its bulk form, hosts a unique $4a\times4a\times3.2c$ (where $a$ and $c$ are the in-plane and out-of-plane lattice constants, respectively) charge density wave (CDW) below $T_c=105~K$ \cite{Tsutsumi1981,Tsutsumi1982,Bayard1976}. This CDW can be altered by V self-intercalation \cite{Disalvo_Waszczak_1981}, by pressure \cite{Friend1978}, and in-plane strain \cite{Zhang_2017}. The CDW transition temperature has been found to depend non-monotonically on thickness \cite{Yang_Wang_2014,Xu_Chen_2013, Pásztor_2017_2dmat}. 

VSe$_2$ is among the most three-dimensional TMDs, with sizable interlayer interaction resulting in significant dispersion in the energy spectrum along k$_z$ \cite{Sato2004,Strocov2012}, making it notoriously difficult to exfoliate. Investigating monolayer (ML) VSe$_2$ has thus relied on molecular beam epitaxy (MBE) \cite{Bonilla_2018, Feng_2018,Liu_2018,Umemoto_2018,Chen_2018,Duvjir_2018,Liu_2019,Wong_2019,Duvjir_2019,Zhang_2019,Chen_2020,Kezilebieke_2020,Chua_2020, Chua_2021, Huang_2022}. Interest in this compound is primarily fueled by the ferromagnetic (FM) order predicted by theory in ML VSe$_2$ \cite{Ma_2012, Lebègue_2013, Fuh_2016}. 
Following the first experimental observation of strong room temperature ferromagnetism in ML VSe$_2$ \cite{Bonilla_2018}, several subsequent studies reported only weak \cite{Duvjir_2018, Kezilebieke_2020, Chen_2020} or no magnetic signal \cite{Chen_2018, Feng_2018, Umemoto_2018} in these MBE-grown samples, independently of the substrates. Theory has shown that the CDW plays a crucial role in the magnetic properties of ML VSe$_2$, with competition from the CDW suppressing ferromagnetic ordering \cite{Fumega_2019, Coelho_2019}. However, observations on MBE-grown ML VSe$_2$ are widely scattered regarding magnetism and charge order. Reported CDW transition temperatures $(T_\text{CDW})$ range from 350~K (ML grown on bilayer graphene on SiC substrate (BLG-SiC)) \cite{Duvjir_2018} to no transition at all (ML grown on NbSe$_2$ substrate) \cite{Kezilebieke_2020}. Such widely varying behavior is even reported for ML grown on the same substrate. For example, very different $T_\text{CDW}$ of 140~K \cite{Feng_2018, Umemoto_2018}, 171~K \cite{Chen_2020}, 220~K \cite{Chen_2020}, and 350~K \cite{Duvjir_2018} have been reported for ML VSe$_2$ on BLG-SiC.

Likewise, the reported energy gap in the quasi-particle spectrum associated with the CDW ground state ranges between 9~meV \cite{Duvjir_2018} and 230~meV \cite{Umemoto_2018} for ML VSe$_2$ on BLG-SiC. The same variability has been observed on other substrates such as highly ordered pyrolytic graphite or MoS$_2$. Finally, there is no consensus on the real-space structure of the CDW either. Although theoretical works support the possibility of different CDW reconstructions subject to strain, doping, and interaction with the substrate \cite{Zishen_2021, Fumega_2023}, the variability of the experimentally observed charge modulations (even on the same substrates) is astonishing: $3.2a\times 2.24a$ \cite{Bonilla_2018},  $4a\times 4a$ \cite{Feng_2018, Umemoto_2018}, $\sqrt{3}a\times 2a$ \cite{Duvjir_2018, Wong_2019}, $\sqrt{3}a\times \sqrt{7}a$ \cite{Duvjir_2018, Liu_2018, Chen_2018, Wong_2019}, $4.2a\times 4.6a$ together with a $2.33a$ stripe phase \cite{Chen_2020}, and two simultaneous stripe phases with $4a$ and $2.8a$ periodicity \cite{Chua_2021} have been reported. 

\begin{figure*}[ht!]
    \centering   \includegraphics[width=2\columnwidth]{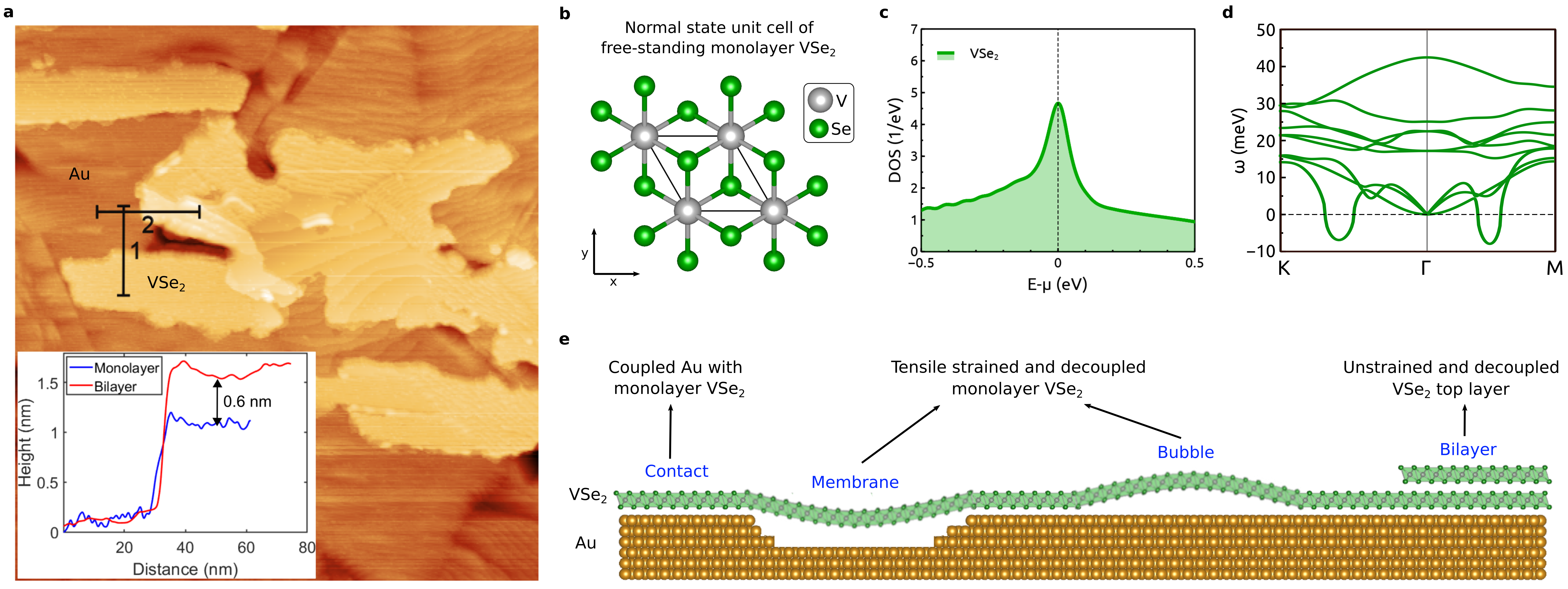}
    \caption{(a) STM topography (380~$\times$~380 nm$^2$, $V_{bias}=500$~mV, $I_t=20$~pA ) of exfoliated ML and BL VSe$_2$ flakes on a Au(111) mosaic substrate. (b) Normal state unit cell of free-standing monolayer VSe$_2$. (c) Density of states (DOS) for free-standing or decoupled monolayer VSe$_2$. (d) Harmonic phonon calculations for free-standing monolayer VSe$_2$. (e) Schematic of the sample with ML and BL flakes of VSe$_2$ (green) on gold (yellow); and possible decoupling scenarios: membrane over a pit in the gold or bubbles due to trapped material. }   \label{fig:schematic_VSe2_on_gold}
\end{figure*}

It has been demonstrated that in TMDs such as NbSe$_2$~\cite{Dreher_2021}, TaS$_2$~\cite{Sanders_2016}, and IrTe$_2$~\cite{asikainen_suppression_2025}, interfacing monolayers with Au(111) is an effective strategy to suppress electronic instabilities like CDWs and superconductivity. The suppression arises from hybridization with the substrate, leading to pseudo-doping and a reduction of the effective coupling constants~\cite{Shao_2019}. Moreover, the interaction strength with the substrate depends on the rotational alignment of the constituents~\cite{Pushkarna_2023, Huang_2025}, potentially offering a new means of tunability.

Here, we investigate monolayer VSe$_2$ exfoliated on Au(111) using scanning tunneling microscopy and spectroscopy (STM/STS), with a particular emphasis on the effect of interface coupling on the CDW. Bilayer (BL) and thicker VSe$_2$ flakes host a $4a\times4a$ CDW. Monolayer flakes host a different $\sqrt{3}a\times \sqrt{7}a$ CDW, but only when they are decoupled from the gold substrate. The CDW is absent in coupled ML, where STM topography reveals a moiré pattern consistent with the twist angle. DFT calculations confirm the formation of different CDW patterns in decoupled ML and BL VSe$_2$ and its suppression when the ML is coupled to the gold substrate.  

\section{Results and Discussion}

We prepared ML flakes by exfoliating bulk crystals onto template-stripped gold substrates \cite{Pushkarna_2023} that consist in a mosaic of (111) oriented grains (see Methods). This process yields a large area predominantly covered with macroscopic ML islands and occasional smaller BL regions revealed by scanning tunneling microscopy in Figure~\ref{fig:schematic_VSe2_on_gold}(a). The exfoliated MLs closely follow the gold steps of the substrate and are identified by their apparent height of 8$\pm4${\AA} relative to the Au(111) surface. The BLs appear consistently higher by 6$\pm0.5${\AA} (Figure~\ref{fig:schematic_VSe2_on_gold}(a) inset), corresponding to the interlayer spacing of bulk VSe$_2$ \cite{Rana_Das_Nandi_2025}. 
From a theoretical point of view, free-standing monolayer 1T-VSe$_2$ in its normal state (Figure~\ref{fig:schematic_VSe2_on_gold}b) is a highly unstable phase at low temperatures. 
The calculated electronic density of states (DOS), shown in Figure.~\ref{fig:schematic_VSe2_on_gold}c, reflects this unstable metallic character of the system, with partially occupied V $d$-orbitals and a van Hove singularity dominating near the Fermi level. 
The combination of the van Hove singularity at the Fermi level and electron-phonon coupling leads to the emergence of a CDW at low temperatures \cite{Fumega_2023,Diego_Subires_2024}.
This can be analyzed through the phonon spectrum in Figure~\ref{fig:schematic_VSe2_on_gold}d, where it can be seen that free-standing monolayer VSe$_2$ displays dynamical instabilities leading to the formation of different CDWs. Specifically, at $3/5\overline{\Gamma K}$ and $1/2\overline{\Gamma M}$ points associated with a $\sqrt{3}a \times\sqrt{7}a$ CDW and a $4a \times 4a$ CDW respectively. Anharmonic calculations have shown that $\sqrt{3}a \times\sqrt{7}a$ CDW dominates in the tensile strain regime, while the $4a \times 4a$ CDW occurs in the unstrained and compressive scenarios  \cite{Fumega_2023}.

The situation changes markedly when the ML is exfoliated on the Au(111) substrate. Our sample preparation offers unique opportunities to assess the influence of the substrate on the VSe$_2$ flakes. 
In our sample (Figure~\ref{fig:schematic_VSe2_on_gold}a), we have identified three different cases that are summarized in Figure~\ref{fig:schematic_VSe2_on_gold}e: i) monolayer VSe$_2$ in contact with gold, ii) bilayers of VSe$_2$ on gold, and iii) free-standing membranes and bubble regions of monolayer VSe$_2$ on gold. We will analyze each of these regimes in the following.

\begin{figure}[th!]
    \centering
    \includegraphics[width=1.05\columnwidth]{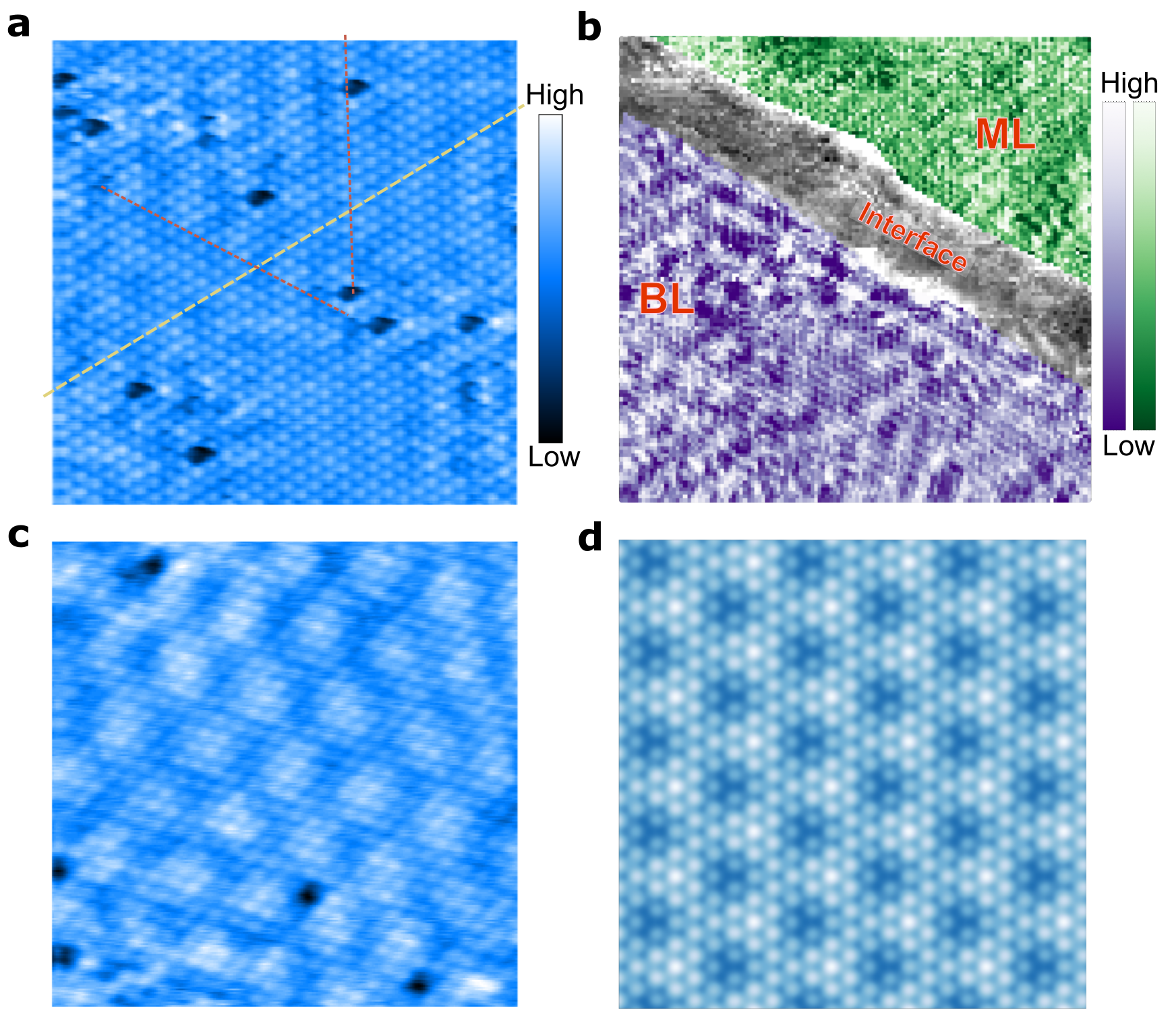}
    \caption{(a) 9 $\times$ 9 nm$^2$ STM topography ($V_{bias}=200$~mV, $I_t=200$~pA ) on a BL region showing a weak $4a\times4a$ CDW modulation along with the atomic lattice. The red and yellow dashed lines mark the CDW and atomic directions, respectively. (b) LDOS map (at $E=-70$~meV) over the boundary (grey) between a BL (purple) and ML (green) region. (c)  8 $\times$ 8 nm$^2$ STM topography ($V_{bias}=-100$~mV, I$_T$ = 200 pA) on bulk VSe$_2$ showing the in-plane $4a\times4a$ CDW modulation along with the atomic lattice. (d) DFT simulated image showing $4a\times4a$ CDW modulation.}

    \label{fig:large_scale_BL}
\end{figure}

\subsection{Monolayer and Bilayer VSe$_2$ on Au(111)}

\begin{figure}[th!]
    \centering
    \includegraphics[width=0.95\columnwidth]{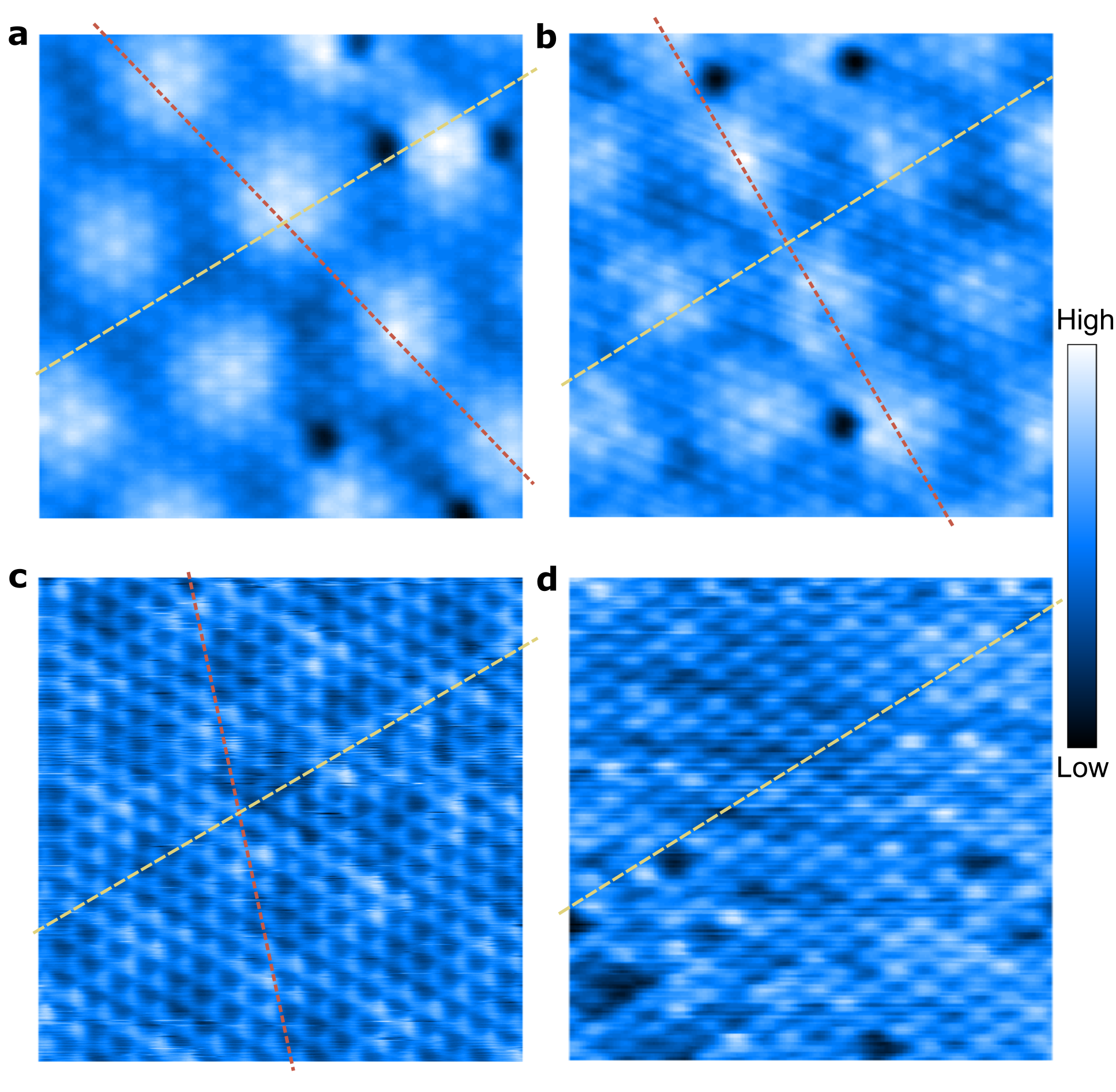}
    \caption{STM topography (5~$\times$~5 nm$^2$, see imaging parameters in Methods) on different ML flakes on Au(111) obtained in a single exfoliation. The red and yellow dashed lines mark the moiré and atomic lattice directions, respectively. Note the same atomic lattice orientation in all four images, as expected for flakes originating from the same single crystal.}

    \label{fig:moire_ML}
\end{figure}

Atomic-resolution STM images of all bilayer flakes reveal a triangular atomic lattice accompanied by a perfectly commensurate, albeit weak, $4a\times4a$ periodic modulation (Figure~\ref{fig:large_scale_BL}(a)), reminiscent of the CDW observed in bulk (Figure~\ref{fig:large_scale_BL}(c)) and thin flakes \cite{Pásztor_2017_2dmat}. When the STM tip is positioned over an adjacent monolayer region, the 
$4a\times4a$ modulation is no longer detected (Figure~\ref{fig:large_scale_BL}(b)). 
Instead, STM topography reveals a variety of regular patterns whose orientation and periodicity along the red-dotted lines in Figure~\ref{fig:moire_ML} vary between flakes located on different gold grains. These patterns are consistent with moiré structures generated between rotated Au(111) grains and MLs obtained from a single crystal exfoliation oriented along the same direction highlighted by the yellow-dotted lines in Figure~\ref{fig:moire_ML}.
At the largest twist angle between VSe$_2$ and Au(111) in Figure~\ref{fig:moire_ML}(d), the moiré pattern is more challenging to identify. 
The assignment of the $4a\times4a$ CDW in the BL flakes and the moiré patterns in the ML flakes is further corroborated by their different bias-dependencies (contrast inversion versus no contrast inversion \cite{Spera_2020}) and by their characteristic behavior near defects (weak pinning of the CDW in VSe$_2$ \cite{Pásztor_2019_PRR, Jolie2019} versus no influence for the moiré), as described in Sections I and II of the Supplementary Material. Note that coupled MLs never show any CDW modulation, although the hybridization strength depends on twist angle.

From a theoretical point of view, the $4a\times4a$ reconstruction observed in the bilayer region is in agreement with DFT simulations of the $4a\times4a$ CDW. 
For the monolayer in contact with gold, ab initio calculation reveal a strong hybridization between VSe$_2$ states and the Au(111) substrate, which modifies the electronic spectrum, as evident from the DOS in Figure~\ref{fig:spectroscopy}c. The interaction with Au(111) induces a pseudo-doping effect \cite{Shao_2019}. 
The hybridization reshapes the electronic bands in such a way that V $d$-states appear as heavily electron-doped relative to the free-standing case. This electronic reconstruction has direct consequences for the lattice dynamics, inducing a stabilization of the normal state and quenching the CDW phases. 
Simulated STM images for the coupled configuration, exhibit a markedly different contrast compared to bilayer case. 
They reproduce the hexagonal lattice symmetry of the monolayer in the normal state, with contrast primarily arising from the Se sublattice. This hexagonal pattern is additionally modulated by the Au(111) substrate, leading to a moiré pattern with an emergent length scale that depends on the angle between Au(111) and monolayer VSe$_2$ (Figure~\ref{fig:decoupled_topography}c).

\subsection{Membranes and Bubbles of Monolayer VSe$_2$ on Au(111)}

\begin{figure}[th!]
    \centering
    \includegraphics[width=1\columnwidth]{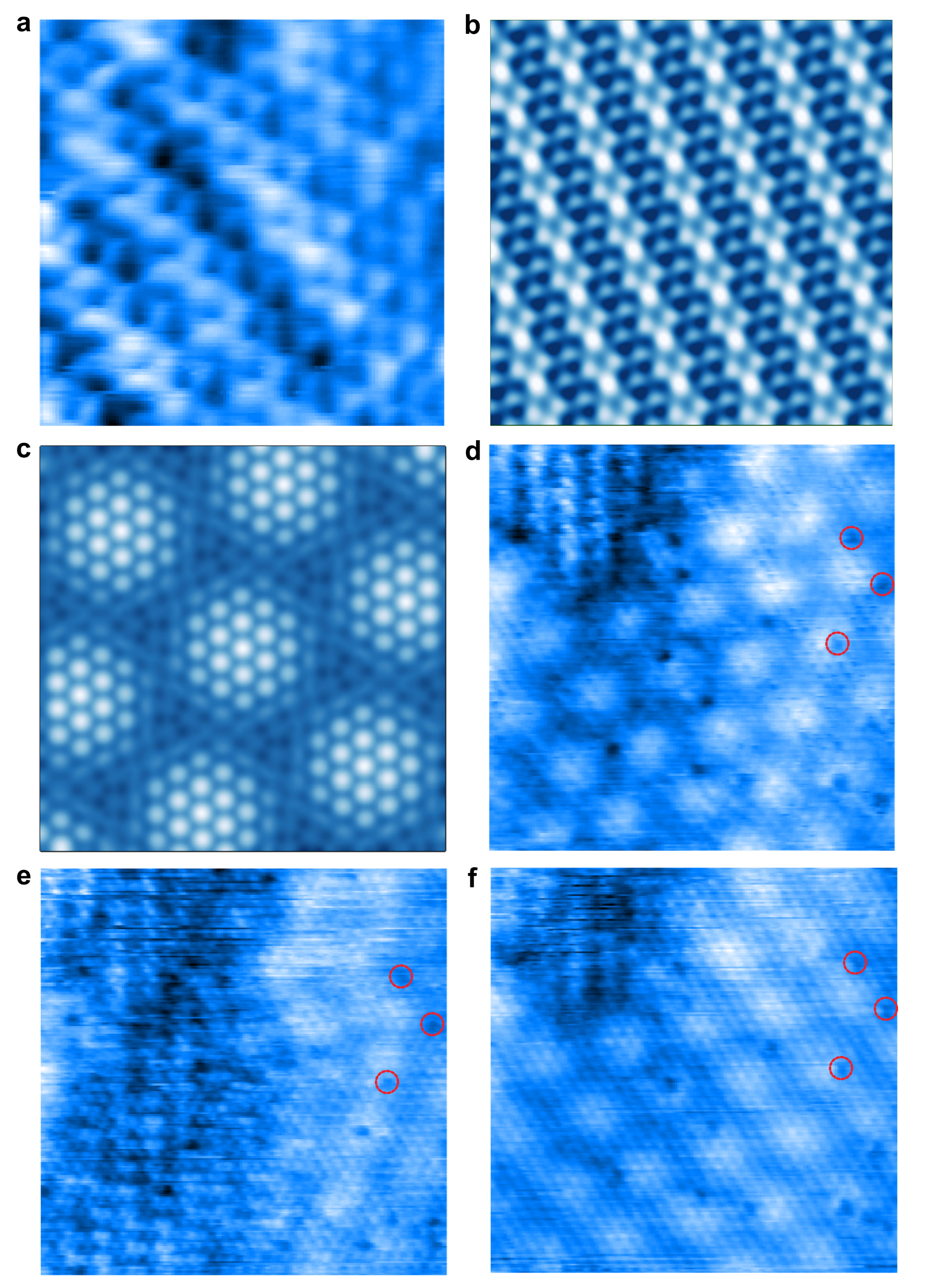}

    \caption{ (a) STM topography of the $\sqrt{3}a \times\sqrt{7}a$ ML-CDW developing on ML bubbles (Image size: 3~$\times$~3 nm$^2$; Setpoint: $V_{bias}=200$~mV, $I_t=50$~pA). (b) Simulated STM images (5~$\times$~5 nm$^2$) of decoupled VSe$_2$ displaying a $\sqrt{3}a \times\sqrt{7}a$ CDW phase and (c) coupled VSe$_2$ showing the absence of CDW but rather a moiré pattern. (d)-(f) Consecutive STM topographic images of the same area of a ML (see the three defects highlighted in red as references). They show a decoupled ML membrane with its ML-CDW surrounded by moiré regions where the ML is coupled to the substrate. It shows the reversible coupling and decoupling of the membrane to the substrate in the lower left region. (10~$\times$~10 nm$^2$; $-200$~mV, $1$~nA).  
        }
    \label{fig:decoupled_topography}
\end{figure}

Decoupled MLs shown in Figure~\ref{fig:decoupled_topography} appear strikingly different from the MLs coupled to gold shown in Figure~\ref{fig:moire_ML}. Both membranes and bubbles reveal a $\sqrt{3}a\times \sqrt{7}a$ periodic modulation in topographic STM images (Figure~\ref{fig:decoupled_topography}(a)), consistent with the monolayer CDW (ML-CDW) previously reported for MLs \cite{Duvjir_2018, Liu_2018, Chen_2018, Wong_2019} and with our DFT calculations for this CDW phase in Figure~\ref{fig:decoupled_topography}(b).

Figures~\ref{fig:decoupled_topography}(d)-(f) are particularly revealing of the effect of coupling a ML to the gold substrate. They show a decoupled membrane with its $\sqrt{3}a\times \sqrt{7}a$ CDW surrounded by the moiré pattern characteristic of a coupled ML. When repeatedly scanning the same area using a high current setpoint where the tip/sample interaction is strong, we observe a reversible switching between the moiré and the ML-CDW in the lower left-hand region of the image. We interpret this as a region where the ML alternates between coupled and decoupled states, providing direct experimental evidence that the interaction with the substrate is responsible for quenching the CDW phase.
 
\begin{figure}[h!]
    \centering
    \includegraphics[width=1\columnwidth]{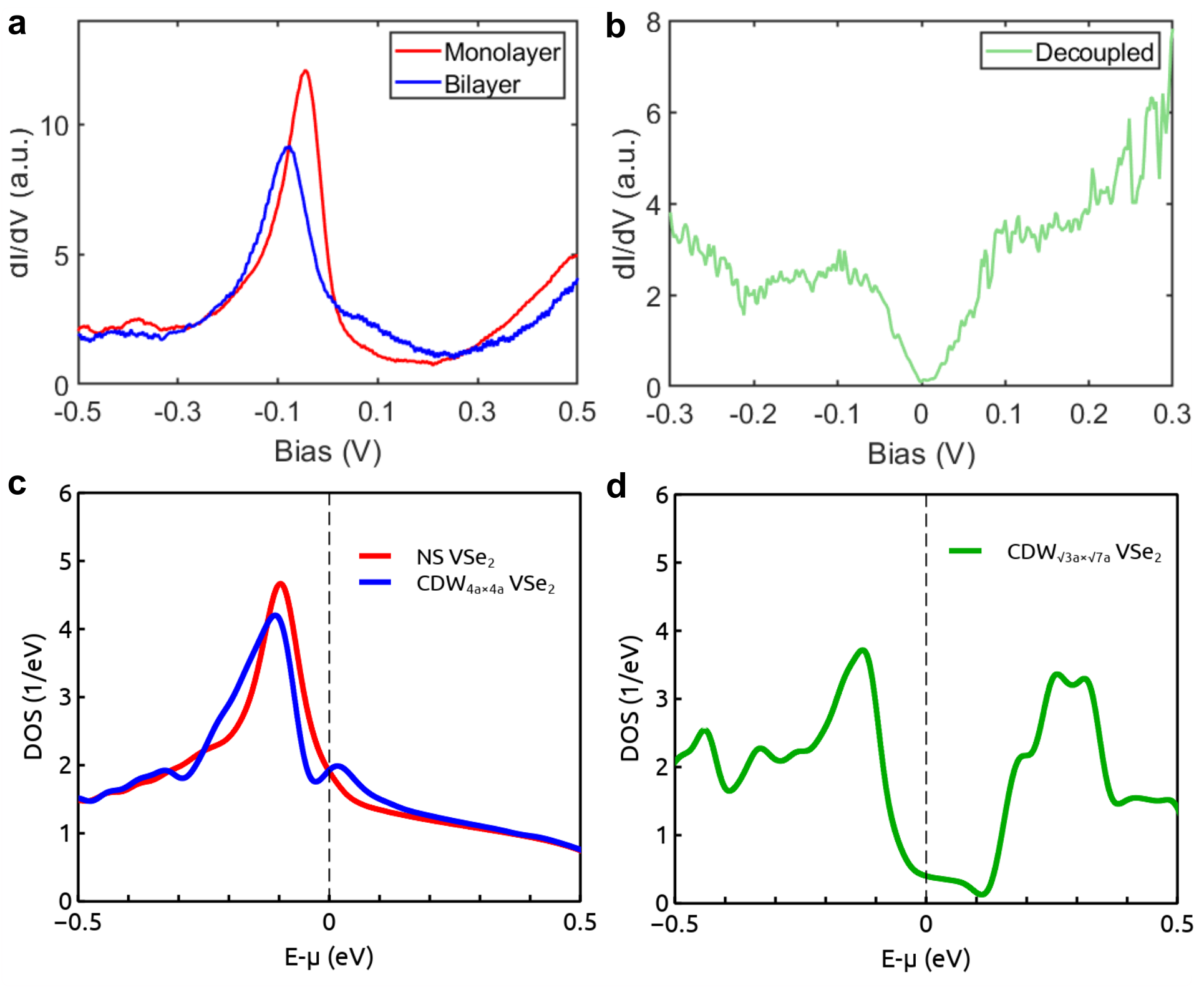}
    \caption{(a) Differential tunneling conductance spectra measured on a coupled monolayer with a moiré pattern and a bilayer with a $4a \times 4a$ CDW. (b) Differential tunneling conductance spectra measured on a decoupled monolayer with a $\sqrt{3}a\times\sqrt{7}a$ CDW. (c) and (d) show the corresponding DFT model calculations. 
        }
    \label{fig:spectroscopy}
\end{figure}

Coupled and decoupled monolayers are also markedly different from a spectroscopic point of view. Coupled ML and BL areas consistently show a metallic spectrum with a strong peak below the Fermi level ($V_b=0$~V). The precise position of the peak depends on thickness, with a shift to lower energies in the BL regions (Figure~\ref{fig:spectroscopy}(a)). These observations are very well reproduced by our DFT calculated DOS in Figure~\ref{fig:spectroscopy}(c).

Decoupled areas with the ML-CDW show more variability than coupled ML in their spectroscopic signatures, ranging from a slightly suppressed peak near the Fermi level to an entirely gapped spectrum, as shown in Figure~\ref{fig:spectroscopy}(b). The latter is consistent with our calculations for a decoupled ML as shown in Figure~\ref{fig:spectroscopy}(d). Bubbles formed during the exfoliation process and membranes suspended over pits in the gold film have been observed in other TMDs and were also found to be electronically different from the coupled ML \cite{Peto2019, Krane2016, Pushkarna_2023}. 

\subsection{Discussion}

Exfoliated VSe$_2$ flakes has been reported to maintain a $4a \times 4a$ CDW modulation down to three layers \cite{Pásztor_2017_2dmat}, however with a transition temperature depending on thickness. The CDW first weakens in thin layers, but strengthens again in the thinnest layers, as a result of dimensional crossover and confinement effects. Our observations for the BL flakes are consistent with these observations, as they still exhibit the $4a \times 4a$ CDW modulation when exfoliated on gold. The bottom layer acts as a decoupling layer for the top layer of BL VSe$_2$. This enables the formation of the $4a \times 4a$ CDW in the top layer. This is in agreement with previous observations reporting that BL VSe$_2$ exhibits a $4a \times 4a$ CDW modulation when grown on less metallic substrates as bilayer graphene \cite{Chen_2020} and graphite \cite{Chua_2021}. 

In our exfoliated ML flakes on Au, we access a substrate-dominated regime where interfacial hybridization and charge transfer become key control parameters of the CDW. In ML VSe$_2$ flakes coupled to gold, the interaction with the substrate completely dominates the CDW behaviour and the CDW is entirely suppressed, as revealed by our experiments and calculations. In addition, ML VSe$_2$ on gold reveals pronounced metallic behaviour, characterised by a strong peak in the DOS near the Fermi level in perfect agreement with theory.
When the coupling to the substrate is removed, as in the case of strained bubbles and suspended membrane-like flakes, the $\sqrt{3}a\times \sqrt{7}a$ CDW emerges, consistent with the absence of hybridization for free-standing ML VSe$_2$ and a tensile regime.

As previously discussed, the suppression of CDW may lead to the emergence of magnetic ordering in exfoliated ML VSe$_2$ flakes on gold. The non-spin-polarized calculations presented in this work reproduce all the main features found in our experiments. However, they do not exclude the emergence of magnetism in the ML limit, specifically at the boundary between coupled and decoupled regions with the gold substrate. 
Since the STM measurements performed here do not provide direct insight into the spin texture of the flakes, another experimental probe is needed to assess the magnetic properties. Optical means (e.g magneto-optical Kerr rotation) to probe these 2D layers are not suitable due to the lack of an in-plane electric field component on the metallic substrate. Other scanning probe techniques, such as magnetic force microscopy (MFM) and scanning SQUID microscopy (SSM), are promising alternatives. Our experimental setup does not provide access to these techniques in the same UHV system as the STM. Therefore, the samples must be transferred, and their surface must be protected. Due to these technical challenges, our first attempts with these techniques do not yet allow us to firmly conclude on the magnetic properties, and they call for further experimental efforts to probe the possible magnetic texture at the interface between ML VSe$_2$ and the gold. 

\section{Conclusion}
  
In conclusion, we investigated both ML and BL VSe$_2$ flakes on a gold substrate, accompanied by theoretical modelling. Our results show that BLs largely retain their bulk-like character and exhibit a $4a \times 4a$ CDW order, whereas MLs are strongly coupled to the substrate, displaying only a moiré supermodulation without CDW formation. When MLs become decoupled from the gold surface, in strained bubbles and suspended membranes, the $\sqrt{3}a \times \sqrt{7}a$ CDW modulation emerges, highlighting the decisive role of interfacial hybridization and charge transfer in suppressing CDW order. Altogether, our findings identify exfoliated VSe$_2$ on gold as a promising platform for investigating the subtle interplay between competing charge-order phases in two dimensions through interface hybridization.

\section{Methods}

\noindent\textbf{Sample preparation}

Commercial VSe$_2$ single crystals (HQ Graphene) were exfoliated using a slightly modified version of the gold-assisted exfoliation protocol described in Ref. \cite{Pushkarna_2023}. Instead of performing the entire exfoliation in a glove box, we placed a large freshly cleaved bulk VSe$_2$ crystal on a freshly exposed template-stripped gold substrate \cite{Pushkarna_2023}, performing the final exfoliation inside the UHV system using the wheel setup described in Ref. \cite{Pásztor_2017_RSI}. This procedure avoids exposing the surface to atmosphere. No further \textit{in-situ} treatment of the sample was done. 

~

\noindent\textbf{STM/STS characterization} 

The scanning tunneling microscopy (STM) and spectroscopy (STS) experiments were performed with a Specs JT Tyto STM at 4.5~K, at a base pressure better than 1$\times10^{-10}$ mBar. We used electrochemically etched tungsten or iridium tips, which were cleaned \textit{in-situ} by Ar$^+$ ion sputtering before conditioning and characterizing on a Au(111) single crystal. STM topographic images were recorded in constant current mode. The $dI/dV(V)$ conductance spectra were acquired using a lock-in amplifier with a sample bias modulation amplitude in the range of 2 to 10~mV at 347 Hz. The imaging bias and current setting for the scan in Figure~\ref{fig:moire_ML} are (a) $V_{bias}=-100$~mV, $I_t=200$~pA, (b) $V_{bias}=300$~mV, $I_t=200$~pA, (c) $V_{bias}=500$~mV, $I_t=100$~pA, (d) $V_{bias}=150$~mV, $I_t=500$~pA.

~

\noindent\textbf{DFT and numerical calculations}

We have performed \textit{ab initio} electronic structure calculations based on density functional theory (DFT)\cite{HK,KS} for monolayer VSe$_2$ in both the free-standing and Au(111)-coupled configurations. Calculations were carried out using the plane-wave pseudopotential method as implemented in the \textsc{Quantum ESPRESSO} package \cite{0953-8984-21-39-395502,0953-8984-29-46-465901}. The generalized gradient approximation in the Perdew–Burke–Ernzerhof (GGA-PBE) scheme was employed for the exchange–correlation functional \cite{PhysRevLett.77.3865}. Core–valence interactions were described using standard ultrasoft pseudopotentials from the PSLibrary \cite{DALCORSO2014337}.  
The kinetic energy cutoff for the plane-wave basis was set to 60~Ry for the wavefunctions and 700~Ry for the charge density. Brillouin zone integrations were performed using a $12 \times 12 \times 1$ Monkhorst–Pack $k$-point mesh. A vacuum spacing of 20~\AA\ was included along the out-of-plane direction to avoid spurious interactions between periodic replicas. 
For the coupled configuration, the VSe$_2$ monolayer was placed on a commensurate Au(111) slab constructed from the experimental lattice constant of Au. 
Importantly, a minimum thickness of six gold layers in the (111) direction was required to capture the spectral behavior observed in the experiments accurately.
To reduce computational cost while maintaining accuracy, the bottom layers of the Au slab were kept fixed during relaxation, while all V and Se atoms and the topmost Au layers were allowed to relax.  
Atomic structures were relaxed until the forces on each atom were smaller than $10^{-4}$~Ry/Bohr, and the total energy convergence threshold was set to $10^{-9}$~Ry.  

Harmonic phonon spectra were obtained using density functional perturbation theory (DFPT), as implemented in \textsc{Quantum ESPRESSO}. The dynamical matrices were computed on a uniform $8 \times 8 \times 1$ $q$-point grid and subsequently Fourier-interpolated to obtain the phonon band structures.  

\section{Acknowledgements}
We thank A. Guipet for technical support with the scanning probes. We thank M. Poggio, P. Karnatak, and K. Kress for collaborating in an effort to detect magnetism in the ML. CR acknowledges support by the Swiss National Science Foundation Grant 10000496 and 227570. The work by AOF and GL was supported by the  Academy of Finland Project No. 369367. We acknowledge the computational resources provided by the Aalto Science-IT project.
\section*{}
\noindent\textsuperscript{*}These authors contributed equally to this work.
 
\bibliographystyle{ieeetr}
\bibliography{ref}

\end{document}


\graphicspath{{figures/}}

\title{Supplementary Material for Tuning competing electronic phases in monolayer VSe\texorpdfstring{$_2$}{2} via interface hybridization}

\author{Ishita Pushkarna\textsuperscript{*}}
\author{Árpád Pásztor\textsuperscript{*}}
\affiliation{Department of Quantum Matter Physics, University of Geneva, 1211 Geneva, Switzerland}
\author{Greta Lupi}
\author{Adolfo O. Fumega}
\affiliation{Aalto University, Department of Applied Physics, 00076 Aalto, Finland}
\author{Christoph Renner}
\affiliation{Department of Quantum Matter Physics, University of Geneva, 1211 Geneva, Switzerland}


\keywords{charge density waves, interface hybridization, VSe$_2$ on gold, heterostructure, moiré, ab initio methods}
\maketitle

\section{Bias-dependent STM imaging of the moiré pattern}
\begin{figure}[!h]
    \centering
    \includegraphics[width=0.75\linewidth]{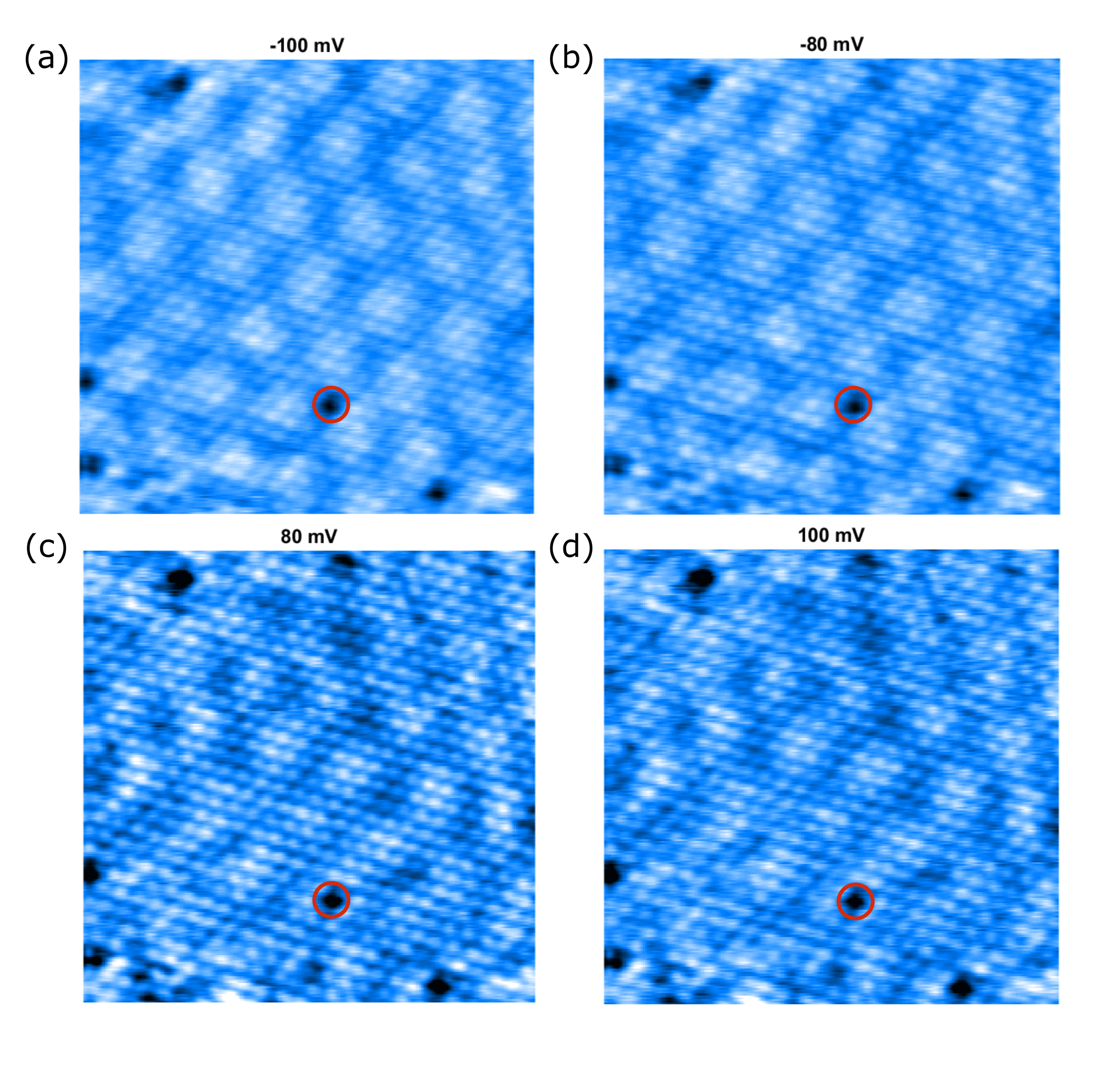}
    \caption{Bias-dependent CDW imaging of the same $ 8 \times 8$ nm$^2$ area on cleaved bulk VSe$_2$ at 4.5~K. Set points are: (a) V$_B$ = -100 mV, I$_T$ = 200 pA, (b) V$_B$ = -80 mV, I$_T$ = 200 pA, (c) V$_B$ = 80 mV, I$_T$ = 200 pA, (d) V$_B$ = 100 mV, I$_T$ = 200 pA (at 4.5 K temperature). CDW contrast is inverting when going from negative to positive bias polarity.}
    \label{fig:vse2_bulk2}
\end{figure}

A decisive confirmation that the observed superstructure originates from moiré patterns rather than a CDW is provided by bias-dependent STM imaging. For a comparison between the superstructure observed in ML and bulk VSe$_2$, we characterized a cleaved-bulk VSe$_2$ crystal, which is reported to show a $4a \times 4a$ CDW modulation \cite{pasztor_dimensional_2017}. As shown in Figure~\ref{fig:vse2_bulk2}, the CDW modulation exhibits a characteristic contrast inversion when the bias is swept across the CDW gap. This can be visualized from the defect (marked by a red circle) at the bottom of the images, which lies in the minima of CDW in panels (a) and (b) when the imaging bias is negative. However, in panels (c) and (d), CDW contrast inverts with respect to the negative bias, and now the defect lies on the maxima of CDW periodicity. This variation of contrast is a characteristic of CDW order, and is reported several times in the literature \cite{pasztor_multiband_2021, Spera_2020}.

\begin{figure}[!h]
    \centering
    \includegraphics[width=0.95\linewidth]{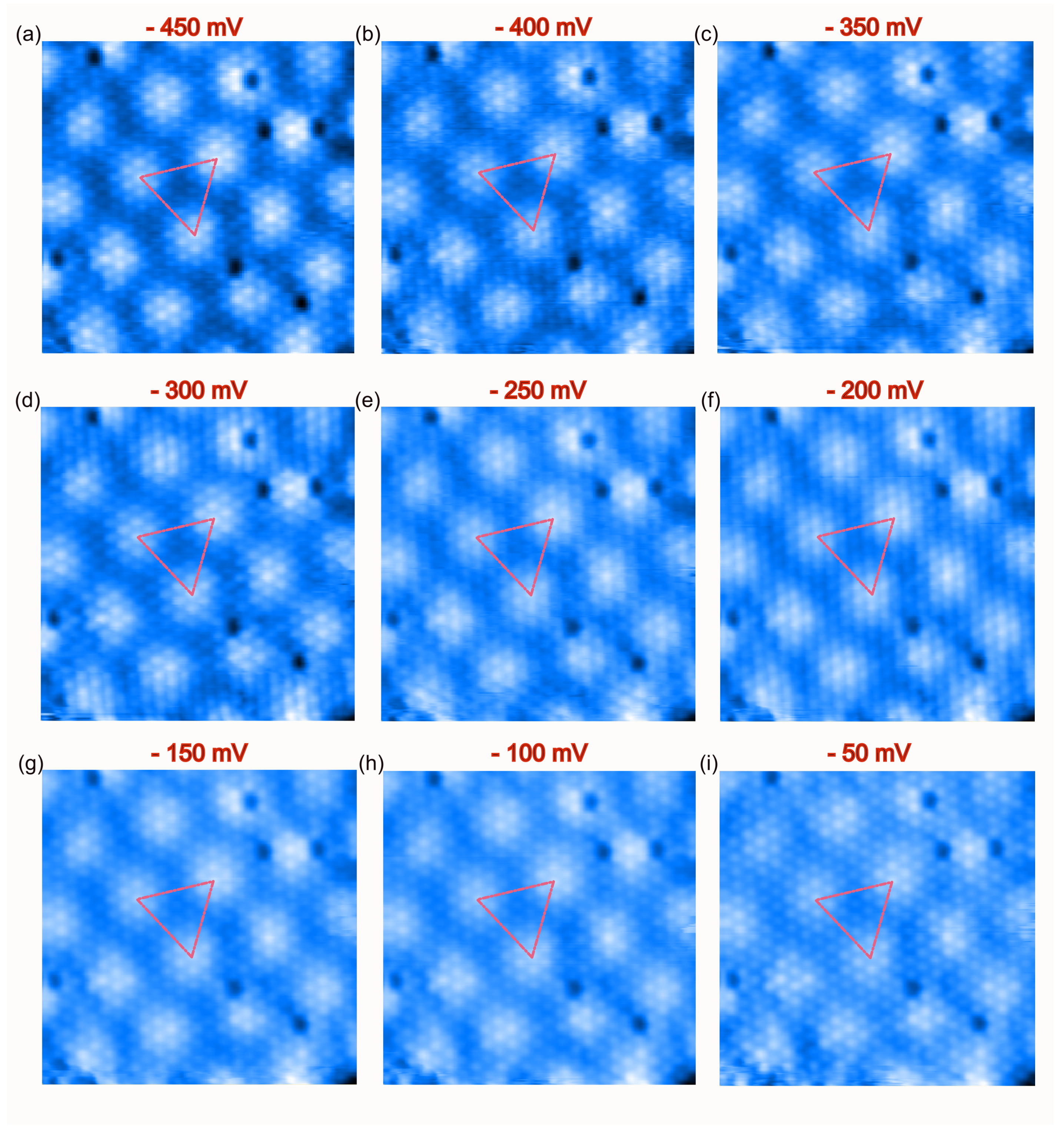}
    \caption{Bias-dependent $7 \times 7$~nm$^2$ STM images from -450~mV to -50~mV of the moiré superstructure observed on ML VSe$_2$ on Au(111). The same overlaid red triangle connects three moiré maxima in all images, confirming that the moiré pattern does not shift as a function of bias.}
    \label{fig:biasdep_negative}
\end{figure}

\begin{figure}[!h]
    \centering
    \includegraphics[width=0.95\linewidth]{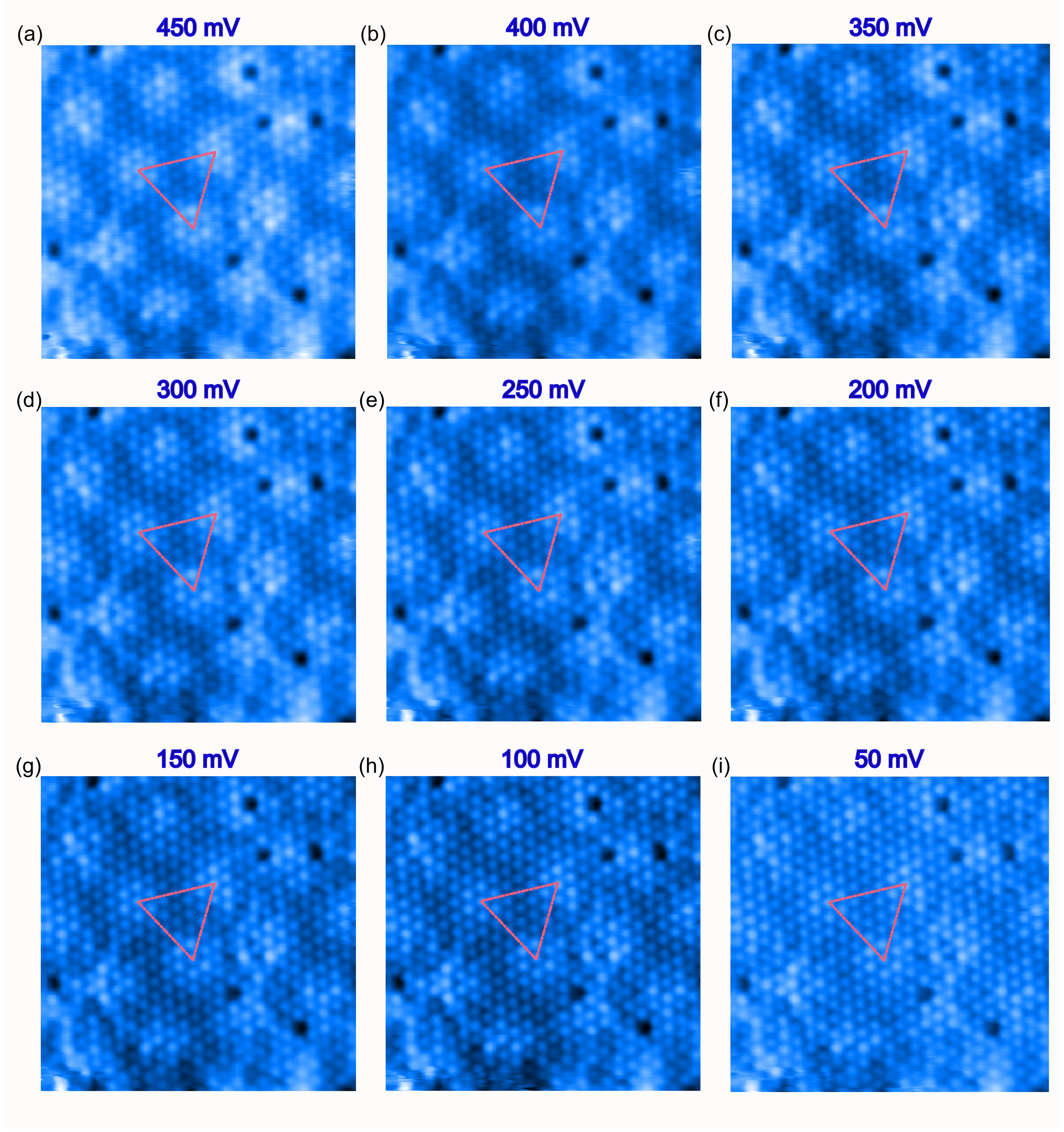}
    \caption{Bias-dependent $7 \times 7$~nm$^2$ STM images from +450~mV to +50~mV of the moiré superstructure observed on ML VSe$_2$ on Au(111). The same overlaid red triangle (same as in Figure~\ref{fig:biasdep_negative}) connects three moiré maxima in all images, confirming that the moiré pattern does not shift as a function of bias and polarity.}
    \label{fig:biasdep_positive}
\end{figure}

In contrast to the bulk VSe$_2$, our measurements on exfoliated ML VSe$_2$ reveal no such inversion for the superstructure. Here, we present a representative dataset, in Figure \ref{fig:biasdep_negative}, showing STM images acquired over the moiré with a twist angle of 2.5°, where both the VSe$_2$ atomic lattice and the superimposed moiré modulation are clearly resolved at different negative biases. All images were obtained over the same scan area, and the uniformity of moiré from –450~mV to –50~mV is verified by the positions of defect sites with respect to the moiré unit cell. Complementary measurements at positive biases, shown in Figure~\ref{fig:biasdep_positive}, span from +50~mV to +450~mV. Here, the moiré contrast (relative to the atomic lattice) gradually diminishes with bias, yet the positions of its maxima and minima remain fixed, consistent across both positive and negative biases.

\clearpage
\newpage
\section{Uniformity of moiré amplitude versus spatial variation of CDW}
It has been reported that the amplitude of each CDW component can vary from one region to another (in the same scan direction), even when there is no occurrence of tip change \cite{Pasztor2019}. This is another clear feature that can be used to distinguish CDWs from moiré patterns. In this dataset, we observe that the amplitude of the moiré pattern remains uniform in all directions, giving rise to a symmetric modulation across spatial locations and over the entire bias range. Moreover, while the CDW shows weak pinning around defect sites \cite{Pasztor2019}, the moiré pattern persists as a continuous and undisturbed modulation, unaffected by defects. Finally, the CDW lattice is aligned and commensurate with the atomic lattice in VSe$_2$ \cite{Pasztor2019}, which is not the case for the moiré pattern. These features further confirm that the modulation observed here is a moiré pattern rather than a CDW.

\bibliographystyle{ieeetr}
\bibliography{ref}